\documentclass[twocolumn,showkeys,superscriptaddress,preprintnumbers,amsmath,amssymb,showpacs,pre]{revtex4-1}
\usepackage{graphicx}
\usepackage{dcolumn}
\usepackage{bm}
\usepackage{amsmath}
\usepackage{color}
\definecolor{Red}{rgb}{1.00, 0.00, 0.00}

\newcommand{\eg}{\textit{e.g. }}
\newcommand{\ud}{\mathrm{d}}

\newcommand{\jb}[1]{\textcolor{red}{#1}}

\begin{document}

\title{Seismic-like organization of avalanches in a driven long-range elastic string as a paradigm of brittle cracks}

\author{Jonathan Bar\'{e}s}
\affiliation{Laboratoire de M\'{e}canique et G\'{e}nie Civil, Universit\'{e} de Montpellier, CNRS, Montpellier, France}
\author{Daniel Bonamy}
\affiliation{SPEC/SPHYNX, DSM/IRAMIS CEA Saclay, Bat.772, F-91191 Gif-sur-Yvette France}
\author{Alberto Rosso}
\affiliation{LPTMS, CNRS, Univ. Paris-Sud, Universit\'e Paris-Saclay, 91405 Orsay, France}


\begin{abstract}
Very often damage and fracture in heterogeneous materials exhibit bursty dynamics made of successive impulse-like events which form characteristic aftershock sequences obeying specific scaling laws initially derived in seismology:  Gutenberg-Richter law, productivity law, B\r{a}th's law and Omori-Utsu law. We show here how these laws naturally arise in the model of the long-range elastic depinning interface used as a paradigm to model crack propagation in heterogeneous media. We unravel the specific conditions required to observe this seismic-like organization in the crack propagation problem. Beyond failure problems, the results extend to a variety of situations described by models of the same universality class: contact line motion in the wetting problem or domain wall motion in dirty ferromagnet, to name a few.
\end{abstract}

\date{\today}


\maketitle

\section{Introduction}

Crackling systems encompasses a broad range of systems; those who, under slowly varying external forcing, respond via series of violent random impulses, so-called avalanches. Crack growth \cite{maloy06_prl,bonamy2009_jpd,bonamy2011_pr,bares14_prl}, damage \cite{alava06_ap,petri1994_prl,baro13_prl,makinen2015_prl,ribeiro2015_prl} or plasticity spreading in a stressed solid \cite{miguel2001_nat,papanikolaou2012_nat,zapperi1997_nat,bares2017_pre}, magnetization change in ferromagnets \cite{urbach95_prl,durin05_book,ledoussal2010_epl}, imbibition of a porous media \cite{ertas1994_pre,rosso2002_pre,planet09_prl,snoeijer2013_arfm}, earthquakes \cite{bak02_prl,corral04_prl,langenbruch2011_grl,davidsen2013_prl}, neuronal activity \cite{beggs2003_jn,bellay2015_elife}, strain in shape-memory alloys \cite{balandraud2015_prb}, magnetic vortex dynamics in superconductor \cite{field1995_prl,altshuler2004_rmp} \textit{etc}, are illustrative examples of cracking noise. A key feature in these systems is that the individual avalanches exhibit universal scale-free statistics and scaling laws, independent of the microscopic and macroscopic details but fully set by generic properties such as symmetries, dimensions and interaction range (see \cite{sethna01_nature} for review). Those are understood in the framework of the depinning transition of elastic manifolds, separating a quiescent phase where the system is trapped by the landscape disorder and an active phase where the applied forcing is sufficient to make the manifold escape from all metastable states and evolve at finite speed \cite{kardar98_pr,fisher98_pr}. Functional Renormalization theory (FRG) then provides the relevant framework to describe the observed features \cite{chauve01_prl,rosso09_prb,dobrinevski2015_epl,thiery2015_jsm,thiery2016_pre}. 

Beyond the specific scale-free features obeyed by individual avalanches, crackling systems sometimes displays temporal correlations, which is \eg manifested by power-law distributed waiting time between successive events \cite{bak02_prl,baro13_prl,ribeiro2015_prl,bares2018_natcom}. Another illustrative example is found in seismology; earthquakes get organized into aftershock ($AS$) sequences which obeys characteristic laws \cite{arcangelis2016_pr}: Productivity law \cite{utsu1971_jfs,helmstetter2003_prl} stating that the number of produced aftershocks goes as a power-law with the mainshock ($MS$) energy; 
B\r{a}th's law \cite{bath1965_tec} stipulating that the ratio between the $MS$ energy and that of its largest $AS$ is independent of the $MS$ magnitude; and Omori-Utsu law \cite{omori94_jcsiut,utsu1972_jfs,utsu95_jpe} telling that the production rate of $AS$ decays algebraically with the elapsed time since $MS$. These laws, referred to as the fundamental laws of seismology, are central in the implementation of probabilistic forecasting models of earthquakes \cite{ogata1988_jasa}. They are not specific to seismology, but were also reported, at the lab scale, in the acoustic emission associated with the damaging of different materials loaded under compression \cite{baro13_prl,makinen2015_prl}, in the global dynamics of a sheared granular material \cite{zadeh2018_arxiv} and in the simpler situation of a single tensile crack slowly driven in artificial rocks \cite{bares2018_natcom}. In the latter case, it has been possible to show that the fundamental laws of seismology are direct consequences of the individual scale free statistics of both the event sizes and inter-event waiting times \cite{bares2018_natcom,bares2018_ptrs}; productivity and B\r{a}th's law \cite{bath1965_tec} for $AS$ sequences result from the power-law distribution of sizes and Omori-Utsu law results from the power-law distribution of waiting time.  

Noticeably, the simplest (and standard) picture of elastic manifolds driven quasistatically in a random potential fails to reproduce the above time clustering features \cite{sanchez2002_prl}. Those can be recovered by adding supplementary ingredients, as \eg memory effects \cite{zapperi1997_nat}, viscoelasticity \cite{jagla2014_prl}, other slow relaxation processes \cite{jagla2010_jgr,papanikolaou2012_nat,aragon2012_pre} or a finite temperature \cite{ferrero2017_prl}. A more general explanation has been proposed in \cite{laurson2009_jsm,font2015njp,janicevic2016_prl}: Power-law distributed inter-event waiting time simply arise when a finite detection threshold is applied to separate the events from the background noise. This argument, together with the power-law distributed sizes and waiting times\jb{,} naturally yield aftershock sequences and seismic laws \cite{bares2018_natcom,bares2018_ptrs}, and that an experimentally finite driving rate naturally implies the use of a finite detection threshold, may provide an explanation of the seismic-like temporal organization widely reported in damage and fracture problems. Still, the specific conditions leading to this organization remains to clarify. 

We report here a theoretical and numerical study of the fracture problem in its most fundamental state: a single propagating crack growing throughout an elastic heterogeneous material. This problem is classically identified with the motion of a one-dimensional (1D) long-range elastic string moving in an effective two-dimensional random media \cite{schmittbuhl1995_prl,ramanathan97_prl,bonamy2008_prl,bares2014_ftp}; the different steps underpinning the description are summarized in section \ref{Sec1}. For some conditions, this motion displays a crackling dynamics, made of successive avalanches obeying the fundamental laws of seismicity (Sec. \ref{Sec2}). The specific conditions required to observe the seismic-like organization of successive events are finally discussed (Sec. \ref{Sec3}).        

\section{Theoretical and simulation framework}\label{Sec1}

\begin{figure}
\centering
\includegraphics[width=\columnwidth]{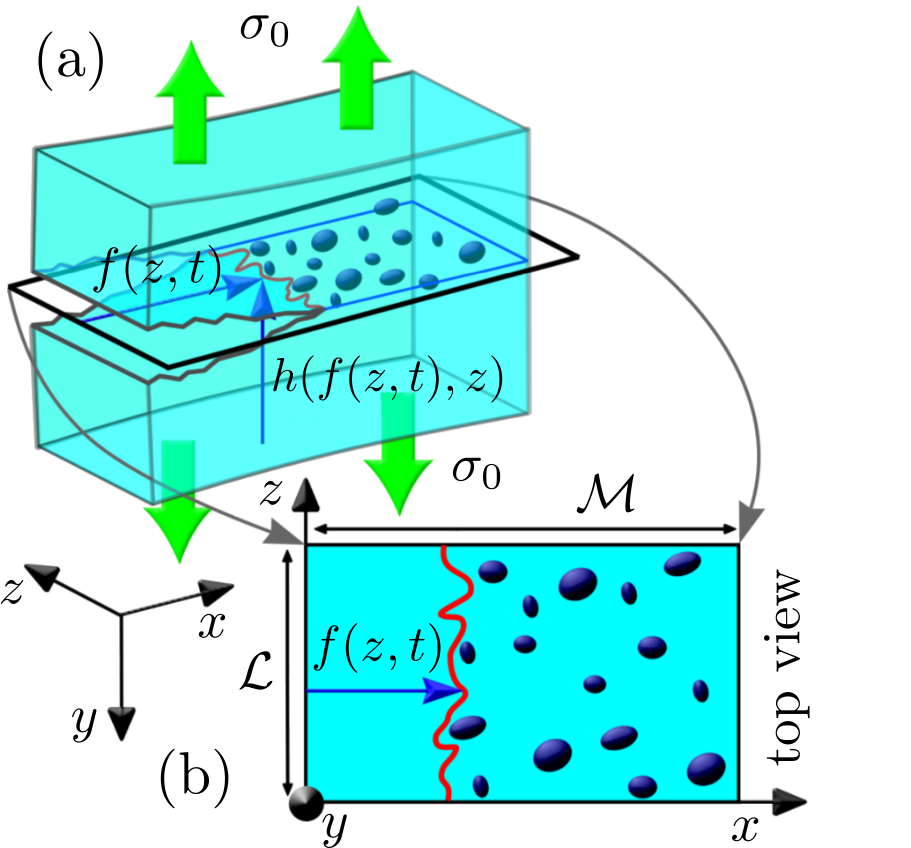}
\caption{Schematic view of a single crack growing in a perfectly brittle heterogeneous material. a: 3D view of the crack propagating from left to right, opened by the stress $\sigma_0$. The crack front shape (red line) is described horizontally by the function $f(z,t)$ and vertically by the function $h(f(z,t),z)$. b: 2D projection on the mean crack plane $(x,z)$. The sample length is $\mathcal{M}$ while its periodic width is $\mathcal{L}$. Ellipses stand for heterogeneities. See the text for more details.}
\label{FigSchemFract}
\end{figure}

The existence of cracks in solids dramatically amplifies applied stresses in their vicinity. This mechanism makes the fracture behavior at the macroscopic scale extremely sensitive to the presence of defects and/or microcracks down to very small scale, which translates into large statistical aspects difficult to assess in practice. For brittle solids under tension, the difficulty is sidetracked by reducing the problem to the destabilization of a {\em single} pre-existing crack in {\em an otherwise intact} material (see \cite{bonamy2017_crp} for a recent review). Strength statistics and its size dependence are analyzed within the Weibull weakest-link framework \cite{Weibull39_prsier}, and Linear Elastic Fracture Mechanics (LEFM) provides the theoretical framework to describe crack destabilization and further growth (see \eg \cite{Lawn93_book}).  

\subsection{Crack growth in homogeneous materials: Continuum fracture mechanics}

Let us consider the situation depicted in Fig. \ref{FigSchemFract}a of a crack front propagating in a brittle solid embedding microstructural inhommegeneities, loaded by applying tensile stresses $\sigma_0$ (or by imposing a displacement field $u_0$) along its external surfaces. In the following, we adopt the usual convention of fracture mechanics and the axes $x$, $y$ and $z$ align with the mean direction of crack propagation, tensile loading, and mean crack front. Moreover $L$ denotes the specimen thickness along $z$. Continuum engineering mechanics simplifies the problem by:

\begin{itemize}
\item[(i)] coarse-graining the solid into an effective linear elastic homogeneous material of Young modulus $E$;
\item[(ii)] considering a straight crack, without any roughness;
\item[(iii)] averaging the behavior along $z$ to reduce the 3D elastic problem to a 2D one.
\end{itemize}   

\noindent The question of when the crack starts growing is then solved by looking at how the total energy evolves with the crack length, $f$. In a perfectly brittle material, this total energy involves two contributions: the potential elastic energy, $\Pi_{\text{pot}}$, stored in the pulled solid and the energy dissipated to create the crack surfaces, $\Pi_{\text{surf}}$. The former decreases with $f$; in the limit of plates with large $x$ and $y$ dimensions, $\Pi_{\text{pot}}(f) \approx \Pi_{\text{pot}}(f=0) - \sigma_0^2 L f^2/E$. The latter increases linearly with $f$: $\Pi_{\text{surf}}=\Gamma L f$ where $\Gamma$ is the fracture energy. When $\sigma_0$ is small, the evolution of the total energy with $f$ is dominated by the increase of $\Pi_{\text{surf}}$ and the crack is stable. when $\sigma_0$ is large, $\Pi_{\text{tot}}$ is dominated by $\Pi_{\text{pot}}$ which decreases with $f$ and, hence, the crack propagates. Griffith introduces the energy release rate, $G$, defined as $G=-(1/L)(\ud \Pi_{\text{pot}}/\ud f)$ which is the amount of energy released as $f$ increases of a unit step and the propagation criterion is:

\begin{equation}
	G > \Gamma,
\label{EqGriffith}	
\end{equation}

\noindent where, in the limit of plates of large $x$ and $y$ dimensions, $G \approx \sigma_0^2 f/E$ and more generally:

\begin{equation}
	G = \frac{\sigma_0^2 f}{E} \times \mathcal{F} (f/L_i,L_j/L_i),
\label{EqRelease}	
\end{equation}

\noindent where $\mathcal{F}(f/L_i,L_j/L_i)$ is a dimensionless function of the various macroscopic lengths $L_i$ invoked to describe the geometry: the specimen dimensions $L_x$ and $L_y$, the position of the crack, of the loading points, \textit{etc}. 

Once the crack starts propagating, an additional contribution due to kinetic energy, $\Pi_{\text{kin}}$, is to be taken into acount in the total system energy. The crack speed, $v=\dot{f}(t)$, is then selected so that the total {\em elastodynamics} energy released as the crack propagates over a unit length exactly balances the fracture energy: $G^{\text{dyn}}(v)=-(1/v)\ud(\Pi_{\text{pot}}+\Pi_{\text{kin}})/\ud t = \Gamma$. Assuming that the specimen is large enough so that the elastic waves emitted by the propagating crack cannot reflect on the boundaries and come back to perturb the crack motion, this equation reduces to \cite{Freund90_book}:

\begin{equation}
	A(v)G = \Gamma \quad \mathrm{with} \quad A(v)\approx 1-\frac{v}{c_R},
\label{EqFreund}	
\end{equation}
 
\noindent where $c_R$ is the Rayleigh wave speed. For a slow enough motion, Eq. \ref{EqFreund} reduces to:

\begin{equation}
	\dfrac{1}{\mu} v = G - \Gamma,	
\label{EqFreundSlow}	
\end{equation}

\noindent where the effective mobility $\mu$ is given by $\mu = c_R/\Gamma$.

It is worth to note that any situation where the solid is loaded by imposing the external stress breaks in a brutal manner. Indeed, $G$ increases with  $f$ (Eq. \ref{EqRelease}). This means that as soon as the crack starts growing, $G$ increases, making $v$ increase, increasing all the $G$, subsequently $v$, \textit{etc.} Conversely, situations involving a loading by a constant {\em displacement} rate, $\dot{u}_0$ may yield stable crack growth. Indeed, $\sigma_0=k(f) u_0(t)$ where the system stiffness $k(f)$ is always decreasing with crack length. Equation \ref{EqRelease} becomes:

\begin{equation}
	G(f,t) = \frac{\dot{u}_0^2 t^2 k(f) f}{E}\times\mathcal{F}(f/L_i,L_j/L_i)
\label{EqReleaseDisplacement0}	
\end{equation}

In some situations, the above expression yields $G$ decreasing with increasing $f$. Then, the crack propagates in a stable manner, so that $G$ remains always close to $\Gamma$. Without loss of generality, we choose a reference time $t_0$ and crack length $f_0$ so that $G(f_0,t_0)=\Gamma$ (right at propagation onset) and look at the crack dynamics in the vicinity of this reference after having shifted the origin: $f \rightarrow f-f_0$ and $t \rightarrow t-t_0$.  Equation \ref{EqFreundSlow} writes:

\begin{equation}
	\frac{1}{\mu}\frac{\ud f}{\ud t} = \dot{G}t-G'f,
\label{EqReleaseDisplacement}	
\end{equation}

\noindent where $\dot{G}=\partial G/\partial t|_{\{t_0,f_0\}}$ (driving rate) and $G'=-\partial G/\partial f|_{\{t_0,f_0\}}$ (unloading factor) are positive constants. In this stable configuration, the crack first displays a transient, and then grows at a constant speed $v=\dot{G}/G'$.
 
\subsection{Crack growth in heterogeneous materials: Depinning line model of cracks}

Equation \ref{EqReleaseDisplacement} predicts continuous dynamics in stable crack growth situations, in contradiction with the crackling dynamics sometimes observed in experiments \cite{maloy06_prl,bares14_prl}. The depinning approach \cite{schmittbuhl1995_prl,bonamy2008_prl,bares2014_ftp} consists in taking into account the microstructure inhomogeneities by adding a stochastic term in the local fracture energy: $\Gamma(x,y,z)=\overline{\Gamma}+\gamma(x,y,z)$. This induces in-plane ($f(z,t)$) and out-of-plane ($h(f(z,t),t)$) distortions of the front (Fig.\ref{FigSchemFract}a) which, in turn, generate local variations in $G$. To the first order, the variations of $G$ depend on the in-plane front distortion only (Fig.\ref{FigSchemFract}b) and the problem reduces to that of a planar crack ($h(f(z,t),t)=\textrm{const.}$) \cite{movchan1998_ijss}. One can then use Rice's analysis \cite{rice1985_jam,gao1989_jam} to relate the local value $G(z,t)$ of energy release to the front shape, $f(z,t)$ (Fig. \ref{FigSchemFract}b):

\begin{align}
	& G(z,t) =\overline{G}(\overline{f},t)(1+J(z,\lbrace f \rbrace)), \label{EqGaoRice}\\
	& \mathrm{with} \, J(z,\lbrace f \rbrace)=\dfrac{1}{\pi} \times PV\int_{\text{crack front} f}{\dfrac{f(\zeta,t)-f(z,t)}{(\zeta-z)^2}d\zeta}, \nonumber  	
\end{align}

\noindent where $PV$ denotes the principal part of the integral;  the long-range kernel $J$ is more conveniently defined by its $z$-Fourier transform $\hat{J}(q)=-\vert q \vert \hat{f}$. $\overline{G}(\overline{f},t)$ denotes the energy release rate that would have been used in the standard continuum picture, after having coarse-grained the microstructure disorder and having replaced the distorted front by a straight one at the mean position $\overline{f}(t)$ (averaged over the specimen thickness). The application of Eq. \ref{EqReleaseDisplacement} at each point $z$ of the crack front supplemented by Eq. \ref{EqGaoRice} yields:

\begin{equation}
	\dfrac{1}{\mu} \dfrac{\partial f}{\partial t}= \dot{G}t-G'\overline{f}+\overline{\Gamma}J(z,\lbrace f \rbrace)+\gamma(z,x=f(z,t)), 	
	\label{eq1}
\end{equation}

The random term $\gamma(z,x)$ is characterized by two main quantities, the noise amplitude defined as $\tilde{\Gamma}=\langle \gamma^2(x,z)\rangle^{1/2}_{x,z}$ and the spatial correlation length $\ell$ over which the correlation function $C(\vec{r})=\langle \gamma(\vec{r}_0+\vec{r})\gamma(\vec{r}_0)\rangle_{\vec{r}_0}$ decreases \cite{bares2014_ftp}.

Equation \ref{eq1} provides the equation of motion of the crack line. A priori, it involves seven parameters: $\mu$, $\overline{\Gamma}$, $\dot{G}$, $G'$, $\ell$, $\tilde{\Gamma}$ and the specimen thickness $\mathcal{L}$. The introduction of dimensionless time, $t\rightarrow t/(\ell/\mu\overline{\Gamma})$, and space, $\{x,z,f\}\rightarrow \{x/\ell,z/\ell,f/\ell\}$ allows reducing this number of parameter to four. The resulting equation of motion writes:

\begin{equation}
	\dfrac{\partial f}{\partial t}=ct - k \overline{f}+J(z,\lbrace f \rbrace)+\gamma(z,f(z,t)), 
\label{eqLine}	
\end{equation}

\noindent where $c=\dot{G}\ell/\mu\overline{\Gamma}^2$ is the dimensionless {\em loading speed}, $k=G'\ell/\overline{\Gamma}$ is the dimensionless {\em unloading factor}. The two other parameters are the dimensionless system size $N \rightarrow \mathcal{L}/\ell$ and the dimensionless noise amplitude $\tilde{\Gamma} \rightarrow \tilde{\Gamma}/ \overline{\Gamma}$.

\subsection{Numerical methods, avalanche detection and sequence identification}

In the following, both system size and noise amplitude are constant: $N=1024$ and $\tilde{\Gamma}=1$. The line is discretized along $z$: $f(z,t)=f_z(t)$ with $z=1,..., N$ and the time evolution of $f_z(t)$ is obtained by solving Eq. \ref{eqLine} using a fourth order Runge-Kutta scheme, as in \cite{bares13_prl,bares2014_ftp}. The second right hand term in Eq. \ref{eqLine} is obtained using a discrete Fourier transform along $z$ (periodic conditions along $z$). A discrete uncorrelated random Gaussian matrix $\gamma_{z,x}$ is prescribed (zero average and unit variance). The third right-hand term in Eq. \ref{eqLine} is obtained via a linear interpolation of $\gamma_{z,x}$ at $\gamma_{z,x=f_z(t)}$. The parameters $c$ and $k$ in the first right-handed term of Eq. \ref{eqLine} are varied from $10^{-6}$ to $5 \times 10^{-4}$ and from $10^{-4}$ to $0.5$, respectively. The movie provided as a supplementary material illustrates the jerky motion obtained via these simulations. 

The crackling noise signal considered in the following is the instantaneous, spatially-averaged crack speed: 

\begin{equation}
\overline{v}(t)=\frac{1}{N}\sum_{z=1}^{N}\frac{ d f_z}{d  t}.
\end{equation} 

\begin{figure}
\centering \includegraphics[width=0.85\columnwidth]{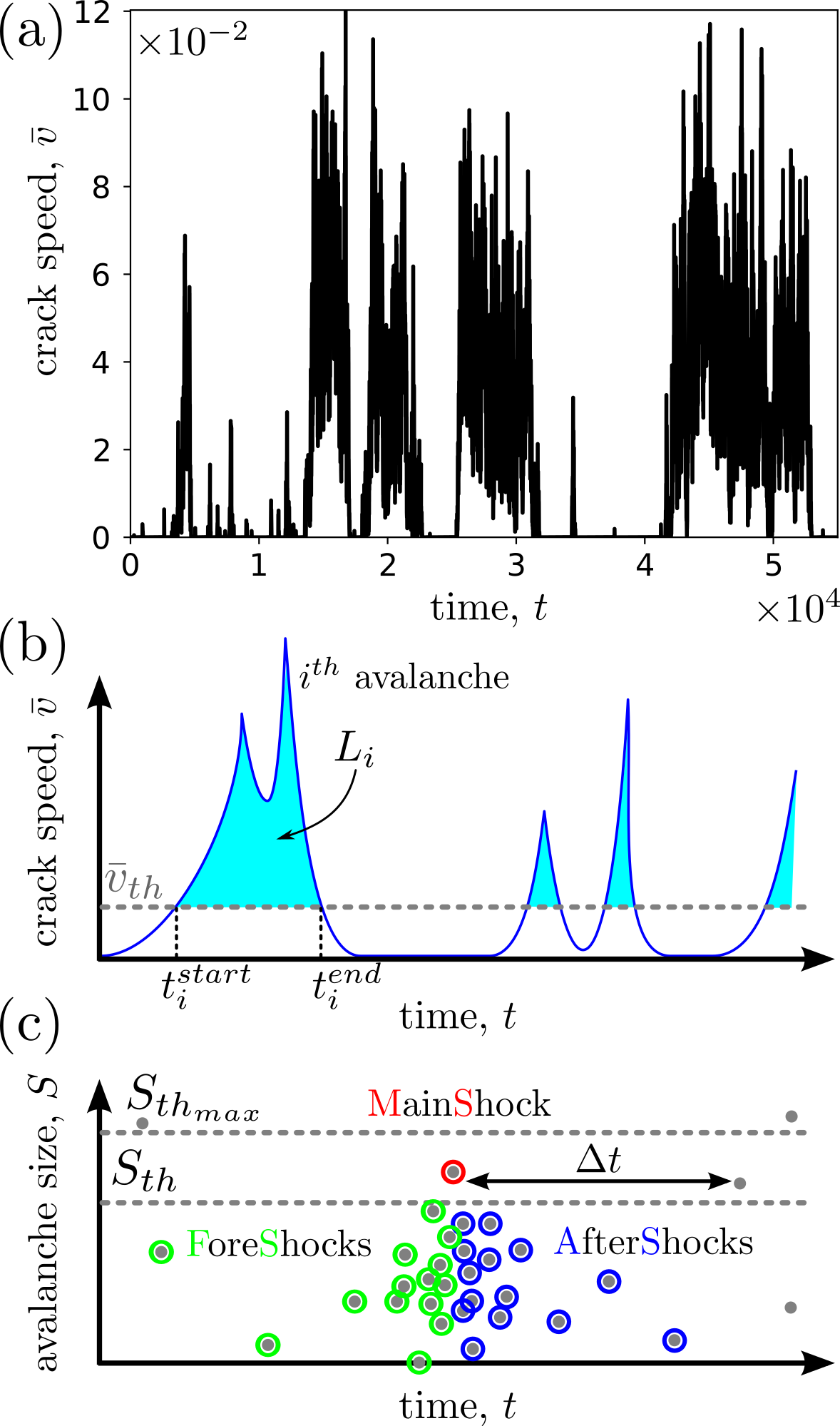}
\caption{
(color online) a: Example of a mean crack speed signal, $\bar{v}(t)$. Here, $c=2 \cdot 10^{-6}$ and $k=1 \cdot 10^{-4}$. Each speed peak corresponds with the crack front jump also called an avalanche. b: Sketch of a mean crack speed signal. Crack speed peak $i$ larger than a threshold $\bar{v}_{th}=c/k$ is detected as an avalanche starting at time $t_i^{start}$ and ending at time $t_i^{end}$. The distance swept by the crack front during this avalanche is the area below the peak, $L_i$ which gives an avalanche size $S_i=N \times L_i$. c: Procedure sketch to identify the $AS$ sequence following a $MS$ (red dot) of size $S_{MS}$ falling within a prescribed range $S_{th}$ to $S_{\text{th}_{\max}} \gtrsim S_{\text{th}}$. The following events until an event of size larger than $S_{MS}$ is encountered are considered as $AS$ (blue points). Along the same line, the preceding events are considered as $FS$ (green points). The waiting time $\Delta t$ is measured between consecutive events larger than a size threshold $S_{th}$.}
\label{fig:v_vs_t}
\end{figure}

\noindent An example of such signal is shown in Fig.\ref{fig:v_vs_t}a. The avalanches are then identified with the bursts of $\overline{v}(t)$ above a prescribed threshold $v_{th}$; an avalanche $i$ starts at $t^{start}_i=t_i$ when the signal rises above $v_{th}$ and ends at $t_i^{end}$ when $\overline{v}(t)$ goes back below this value. The size is then defined by $S_i=N \int_{t_i^{start}}^{t_i^{end}} (\overline{v}(t) - v_{th}) d t$ and the inter-event waiting time between avalanche $i$ and $i+1$ as $\Delta t_i=t_{i+1}-t_{i}$. This is shown in Fig.\ref{fig:v_vs_t}b. In the following, $v_{th}$ has been set to the mean value of $\overline{v}(t)$, denoted as $\langle v \rangle$. Noticeably, $\langle v \rangle = c/k$.   

The so-obtained series of avalanches are finally decomposed into $AS$ sequences. Seismologists have developed powerful declustering methods in this context (see \eg \cite{vanStiphout2012} for a recent review). Most of these methods are based on the spatio-temporal proximity of the events. The spatial proximity is not relevant in this situation with a single crack and, hence, we adopted the procedure proposed in \cite{baro13_prl,makinen2015_prl,ribeiro2015_prl,bares2018_natcom,bares2018_ptrs} and sketched in Fig.\ref{fig:v_vs_t}c:

\begin{itemize}
\item All events with energies in a predefined interval between $S_{th}$ and $S_{th_{max}}$ are considered as $MS$;
\item The $AS$ sequence associated with each $MS$ is made of all events following this $MS$, till an event of size equal or larger than the $MS$ energy, $S_{MS}$, is encountered; 
\end{itemize}

Foreshocks ($FS$) are defined the same way after having reversed the time direction. 

\section{Seismic-like organization of depinning events}\label{Sec2} 

\subsection{Size distribution and Gutenberg-Richter law}\label{distSize}

\begin{figure}
\centering 
\includegraphics[width=0.85\columnwidth]{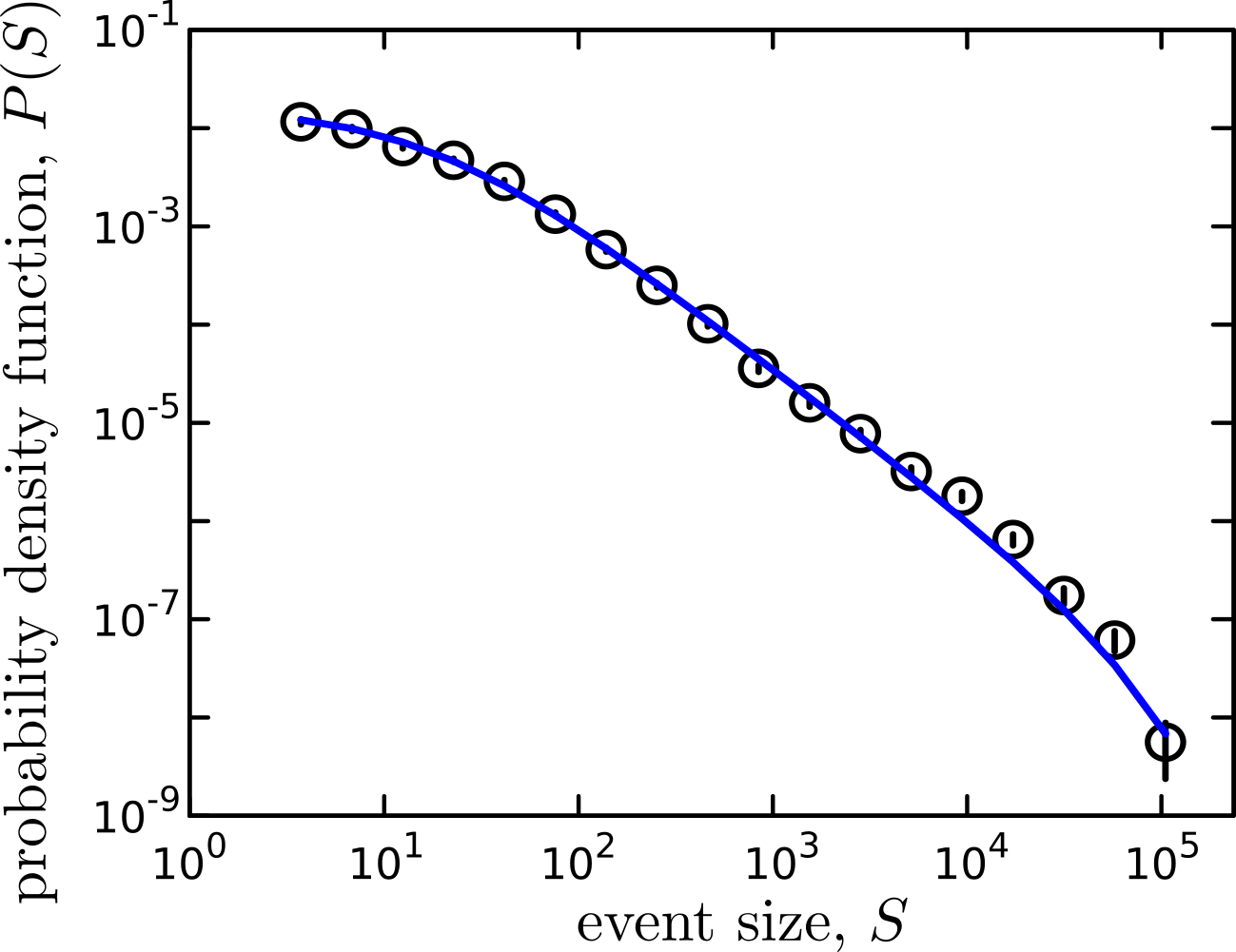}
\caption{(color online) Probability density function of the event sizes $S$ in a simulation with $c=2 \times 10^{-6}$ and $k=10^{-4}$. The axes are logarithmic. The blue plain curve is a fit by Eq. \ref{eqRG}, with exponent $\beta=1.51\pm 0.05$, lower cut-off $S_{min} = 21.3$ and upper cut-off $S_{max}=1.04 \times 10^{5}$.}
\label{fig:GR}
\end{figure}

Figure \ref{fig:GR} shows the probability density function (PDF) to observe an event of size $S$ for a typical simulation. The power-law distribution expected for crackling system is observed over typically $4$ decades. The whole distribution is well fitted by: 

\begin{equation}
	P(S) \sim \frac{e^{-S/S_{max}}}{(1+S/S_{min})^{\beta}}
	\label{eqRG}
\end{equation}

\noindent where $S_{min}$ and $S_{max}$ are the upper and lower cut-offs of the power-law distribution respectively and $\beta$ is the exponent. Both cutoffs depends on the parameters $c$ and $k$. We will return in section \ref{Sec3:phasediagram} to the analysis of these dependencies. Conversely, the size exponent, $\beta = 1.51 \pm 0.05$, barely depends on these parameters (Fig.\ref{fig:GR}), as expected near the depinning critical point of a long range elastic interface within a random potential. Note that the measured exponent is larger than the one expected in the limit of vanishing driving rate: $\beta(c \rightarrow  0) \simeq 1.28$ \cite{bonamy2009_jpd}. As discussed in \cite{jagla2014_prl}, the measure of an apparent, anomalously large Gutenberg-Richter exponent is the signature of avalanche fragmentation in clusters of smaller avalanches strongly correlated in time. 

\subsection{Number of events in $AS$ sequences and productivity law}\label{productivity}

\begin{figure}
\centering 
\includegraphics[width=0.85\columnwidth]{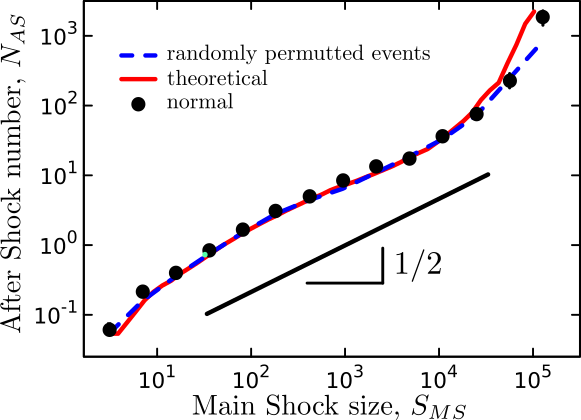}
\caption{(color online) Mean number of $AS$ in the sequence, $N_{AS}$, as a function of the $MS$ size, $S_{MS}$ ($c=2 \times 10^{-6}$ and $k=10^{-4}$). The axes are logarithmic. Black straight line shows an exponent $\alpha=1/2$. Black points are the real data and blue dashed line those obtained after having permuted the sizes and set time step to unity. Plain blue curve is the solution provided by Eq. \ref{EqProd}.}
\label{fig:productivity}
\end{figure}

We now turn to the $AS$ sequences and test whether the scaling laws of seismicity are fulfilled. Figure\ref{fig:productivity} presents the mean number of $AS$, $N_{AS}$, as a function of the size $S_{\text{th}}$ prescribed for the triggering $MS$. In between two cutoffs, $N_{AS}$ goes as a power-law with $S_{th}$ as expected from the productivity law. Following \cite{bares2018_natcom}, we checked that the $N_{AS}$ \textit{v.s.} $S_{MS}$ curve remains unchanged after:
\begin{itemize}
\item having reattributed to each event $i$ the energy of another event $j$ chosen randomly;
\item having arbitrarily set to unity the time interval between to successive events.
\end{itemize} 

This demonstrates that the productivity law is a simple consequence of the size distribution. The relation between the two can be rationalized using the argument provided in \cite{bares2018_natcom,bares2018_ptrs}: The total number of events with a size larger than the prescribed value $S_{MS}$ gives, by definition, the total number of $MS$ of size $S_{MS}$, and hence the total number of $AS$ sequences. The total number of events with a size smaller that $S_{MS}$ gives the total number to be labeled $AS$ in the catalog. The ratio of the latter to the former gives the mean number of $N_{AS}(S_{MS})$. Calling $F(S)$ the cumulative distribution for event size, one gets: 

\begin{equation}
	N_{AS}(S_{MS})=\dfrac{F(S_{MS})}{1-F(S_{MS})}
	\label{EqProd}
\end{equation}

\noindent This equation allows reproducing perfectly the data (plain line in Fig.\ref{fig:productivity}). No fitting parameter are required here. In the scaling regime, $P(S) \sim S^{-\beta}$ with $\beta \approx 1.5$. Hence $F(S) \sim S^{1-\beta}$ and $N_{AS} \sim S_{MS}^\alpha$ with $\alpha=\beta-1\approx 1/2$.

\subsection{Size of the largest aftershock and B\r{a}th law}\label{bath}

\begin{figure}
\centering 
\includegraphics[width=0.85\columnwidth]{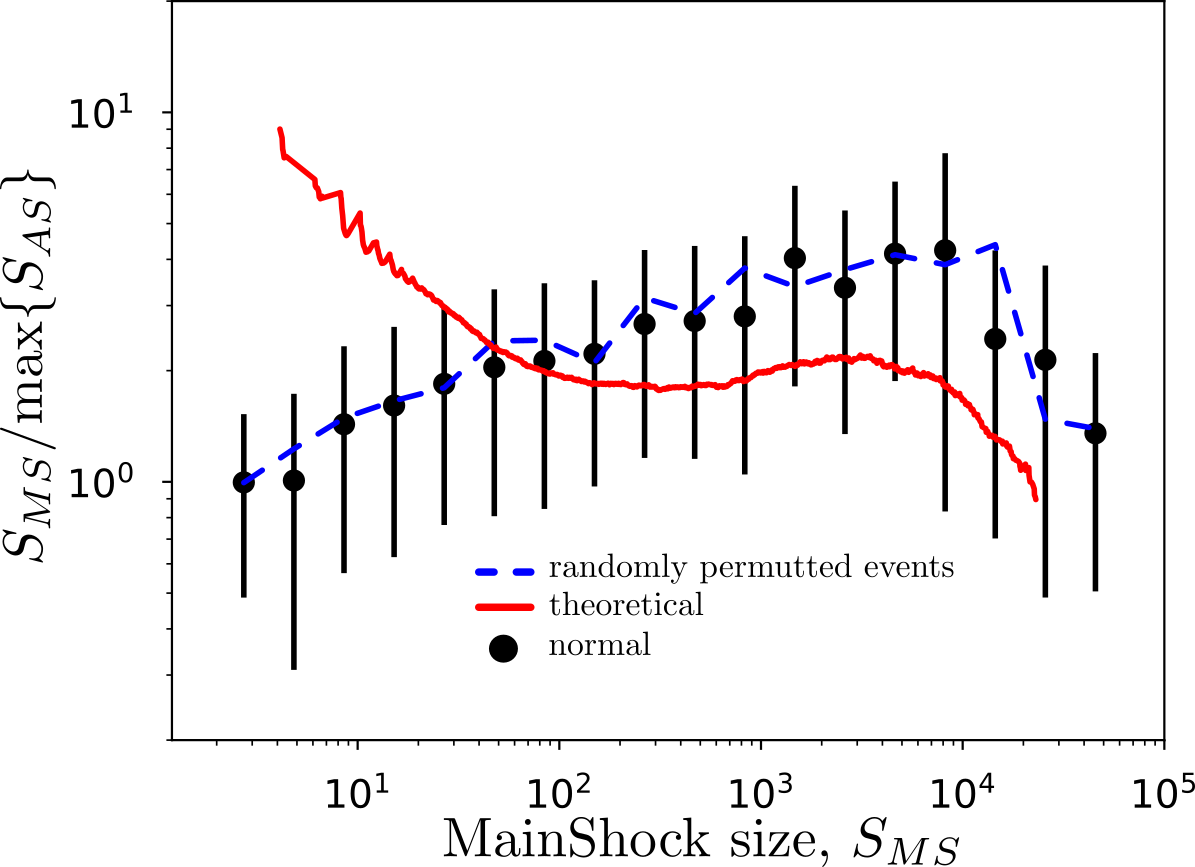}
\caption{(color online) Mean size ratio $\max(S_{MS})/S_{AS}$ between a $MS$ and its largest $AS$, plotted as a function of the $MS$ size, $S_{MS}$ ($c=2 \times 10^{-6}$ and $k=10^{-4}$). The axes are logarithmic. Black points are the real data and blue dashed line those obtained after having permuted the sizes and set time step to unity. Plain red curve is the solutions provided by Eq. \ref{EqBath}.}
\label{fig:Bath}
\end{figure}

The next step is to look at the size ratio between a $MS$ and its largest $AS$. Such a curve is presented in Fig.\ref{fig:Bath}. Once again, permuting randomly the events and setting arbitrarily the time step to unity do not modify the curve. As for the productivity law, this means that this law finds its origin in the size distribution only. Following \cite{bares2018_natcom}, the relation between the two can be derived analytically using extreme value theory (EVT) arguments: Let us call $F_{AS_{max}}(S|N_{AS})$ the probability that the largest $AS$ of a sequence of size $N_{AS}$ is smaller than $S$. All the other $AS$ in the sequence have a size smaller than $S$ so that $F_{AS_{max}}(S|N_{AS})= F(S)^{N_{AS}}$. The mean value $\langle \max (S_{AS}|S_{MS}) \rangle$ of the size of the largest event over the sequences triggered by a $MS$ of size $S_{MS}$ then writes:

\begin{equation}
\begin{split}
	\left\langle \frac{\max (S_{AS})}{S_{MS}} \right\rangle &= N_{AS}(S_{MS}) \\
    &\times \int_{S_{min}}^{S_{MS}} S F(S)^{N_{AS}(S_{MS})-1} P(S) dS
\end{split}
\label{EqBath}
\end{equation}

\noindent where $N_{AS}(S_{MS})$ is given by Eq. \ref{EqProd}. This analytical solution gives a fairly good prediction of the order of $\max (S_{AS})/ S_{MS}$ (see Fig.\ref{fig:Bath}) provided the fact that there is no fitting parameter.

\subsection{Distribution of inter-event time and Bak \textit{et al.} law}\label{distDt} 

\begin{figure}[!h]
\centering \includegraphics[width=0.85\columnwidth]{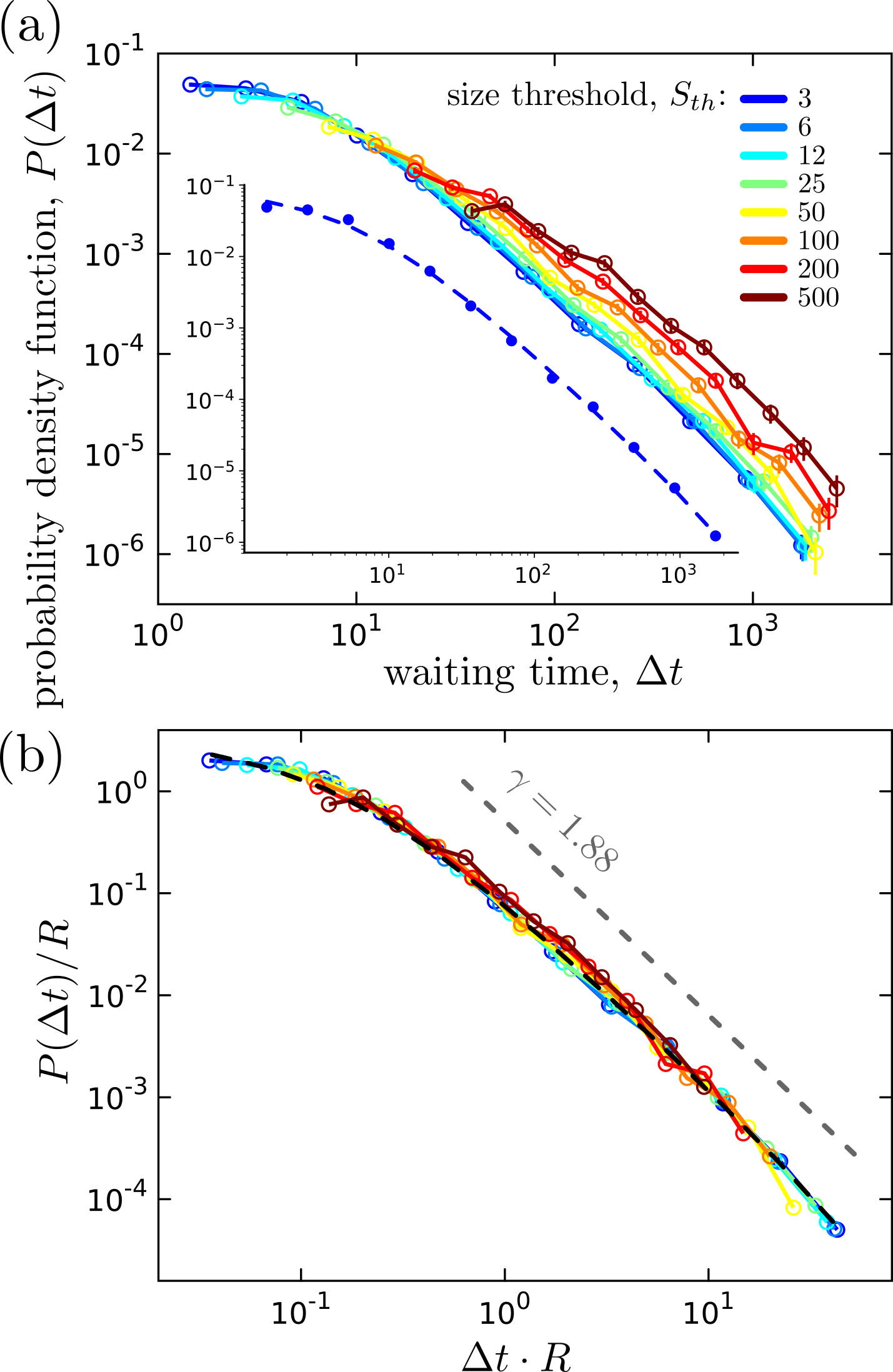}
\caption{(color online) a: Probability density functions of the waiting time $\Delta t$ between two consecutive events of size larger than a prescribed threshold $S_{th}$. Here, $c=1 \times 10^{-5}$ and $k=5 \times 10^{-4}$. The different curves correspond to different values $S_{th}$.  In the inset the dots is the curve reproduced for $S_{th}=3$ and the plain curve is a fit by Eq. \ref{eqWT}, with exponent $\gamma=1.88\pm 0.09$, lower cut-off $\Delta t_{min} = 6.0$ and upper cut-off $\Delta t_{\max}=3.8 \times 10^{3}$. b: Collapse obtained after having rescaled $\Delta t$ with the mean activity rate $R(S_{th})$. Straight dashed line is a power-law of fitted exponent $\gamma=1.88$. Black dashed curve is the rescaled fitted curve of the a inset. In both panels, the axes are logarithmic. Vertical bars stand for $95$\% error-bars.}
\label{fig:distDt}
\end{figure}

We now turn to the analysis of the occurrence time of avalanches. Scale-free statistics is observed for the waiting time separating two successive avalanches; as for avalanche sizes, the whole distribution is well fitted by (Fig.\ref{fig:distDt}a): 

\begin{equation}
	P(\Delta t) \sim \frac{e^{-\Delta t/\Delta t_{max}}}{(1+\Delta t/\Delta t_{min})^\gamma}
	\label{eqWT}
\end{equation}

\noindent where the two time cutoffs $\Delta t_{min}$ and $\Delta t_{max}$ bound the scale free statistics, and $\gamma$ refers to the exponent in between. Same statistics is observed when only the events of size larger than a prescribed threshold, $S_{th}$, are considered (Fig.\ref{fig:distDt}a). The parameters $\gamma$ and $\Delta t_{max}$ barely depend on $S_{th}$. Conversely, the lower cutoff $\Delta_{min}$ increases with $S_{th}$. As observed for seismic events \cite{bak02_prl,corral04_prl} or for AE produced in fracture experiments at lab scale \cite{baro13_prl,stojanova14_prl,makinen2015_prl,ribeiro2015_prl,bares2018_natcom,bares2018_ptrs} and for sheared granular material \cite{zadeh2018_arxiv}, all curves collapse onto a single master curve (Fig. \ref{fig:distDt}c), once time is rescaled with the activity rate $R(S_{th})$, defined as the total number of events divided by the simulation duration: 

\begin{equation}
	P(\Delta t|E_{th}) \sim R(S_{th}) \times f\left(u=R(S_{th})\times \Delta t\right)
	\label{eqWTrescaled}
\end{equation}

\noindent with	$f(u) \sim (1+u/b)^{-\gamma}e^{-u/B}$. The fact that $f(u)$ takes the form of a gamma distribution underpins a stationary statistics for the event series \cite{corral04_prl,ribeiro2015_prl,bares2018_natcom}. The two rescaled time cutoff $b$ and $B$ relates to $\Delta t_{min}$ and $\Delta t_{max}$ via $b=R\times \Delta t_{min}$ and $B=R\times \Delta t_{max}$, where $R$ denotes the mean activity rate during the simulation (total number of avalanches divided by the total duration of the simulation).   
These three parameters $\gamma$, $b$ and $B$ can be interrelated using the conditions $\int_0^{\infty}f(u)du=1$ (normalization of the probability density function $P(\Delta t|E_{th})$) and $\int_0^{\infty}uf(u)du=1$ (since $\langle \Delta t\rangle=1/R(S_{th})$). 

\subsection{Production rate of $AS$ and Omori-Utsu law}\label{Omori} 

\begin{figure}[!h]
\centering \includegraphics[width=0.85\columnwidth]{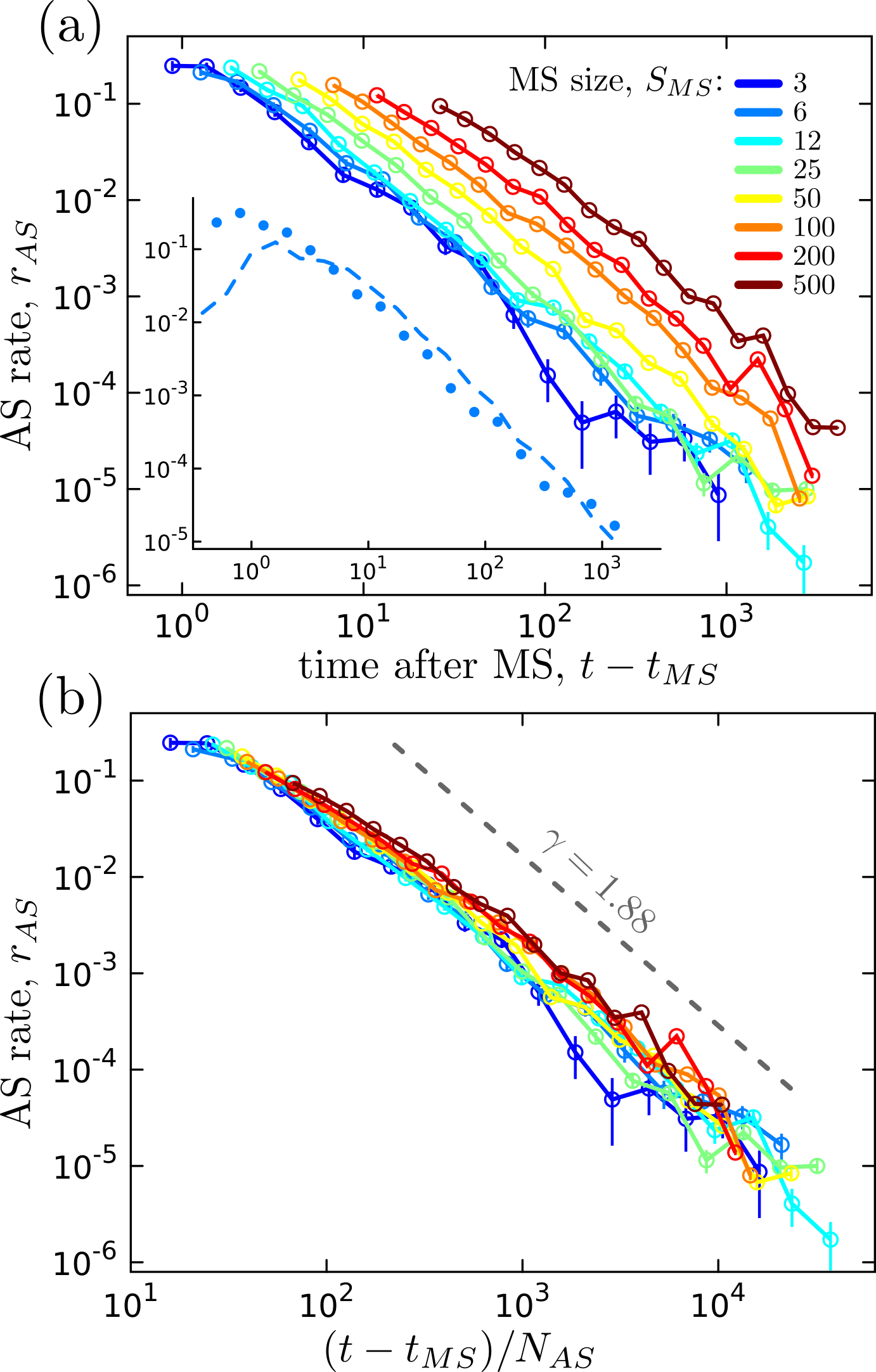}
\caption{(color online) a: Rate of AS, $R_{AS}(t-t_{MS}|S_{MS})$, triggered by a $MS$ of size $S_{MS}$ plotted as a function of the time elapsed since $MS$, $t-t_{MS}$. Here, $c=1 \times 10^{-5}$ and $k=5 \times 10^{-4}$. The different curves correspond to different values of $S_{MS}$. In the inset the dots is the curve reproduced for $S_{th}=6$ and the dashed curve is obtained after having permuted randomly the size $S_i$ attributed to each event occurring at $t_i$. b: Collapse obtained after having set $t-t_{MS} \rightarrow (t-t_{MS})/N_{AS}(S_{MS})$ where $N_{AS}(S_{MS})$ is the mean number of $AS$ produced by a $MS$ of size $S_{MS}$ and is given by Eq. \ref{EqProd}. Straight dashed line is a power-law of exponent $\gamma=1.88 \pm 0.09$ obtained from the analysis of inter-event time (see Fig.\ref{fig:distDt} and Eq. \ref{eqWT}).}
\label{fig:Omori}
\end{figure}

Finally, we looked at the rate of $AS$ produced by a $MS$ of size $S_{MS}$ and its evolution as a function of the time elapsed since $MS$: $R_{AS}(t-t_{MS}|E_{MS})$. To compute these curves, we adopted the procedure developed in \cite{bares2018_natcom}: For each simulation, all sequences triggered by $MS$ of size falling within a prescribed interval are sorted out; subsequently the $AS$ events are binned over $t-t_{MS}$ and the so-obtained curves are finally averaged. Figure \ref{fig:Omori} shows the  resulting curves in a typical simulation. An algebraic decay compatible with the Omori-Utsu law \cite{omori94_jcsiut,utsu95_jpe} is observed (see Fig. \ref{fig:Omori}a) and, within the errorbar, the Omori exponent is equal to the exponent $\gamma$ associated with $P(\Delta t)$ :

\begin{equation}
	R_{AS}(t) \sim \frac{1}{(t-t_{MS})^\gamma}
	\label{eqOmo}
\end{equation}

\noindent As in \cite{bares2018_natcom}, permuting randomly the event sizes in the initial series does not modify the curves observed in Fig. \ref{fig:Omori}. Hence, Omori-Utsu law and the time dependency of $R_{AS}(t|S_{MS})$ find their origin in the scale-free distribution of $P(\Delta t)$, and, hence, the Omori-Utsu exponent is equal to $\gamma$ \cite{bares2018_natcom} and is found to be independent of the $MS$ size $S_{MS}$. Finally, following \cite{bares2018_natcom}, we checked that the dependency with $S_{MS}$ can be fully captured by rescaling $t-t_{MS} \rightarrow (t-t_{MS})/N_{AS}(S_{MS})$ (see Fig. \ref{fig:Omori}b). 

As in \cite{bares2018_natcom}, all curves collapse onto a master curve once $t-t_{MS}$ is rescaled by the mean number of $AS$, $N_{AS}(S_{MS})$, produced by a $MS$ of size $S_{MS}$:
 
\begin{equation}
	R_{AS}(t|S_{MS}) \sim \frac{1}{(1+\frac{t-t_{MS}}{\tau_{min}N_{AS}})^\gamma}
	\label{eqOmoRescaled}
\end{equation}
\noindent The very same relation holds for the $FS$ rate $R_{FS}(t_{MS}-t)$ as the event series are stationary \cite{bares2018_natcom}.

\section{Effect of loading speed and unloading rate}\label{Sec3}

\subsection{On the selection of size distribution}\label{Sec3:RG}

\begin{figure}
\centering \includegraphics[width=1.\columnwidth]{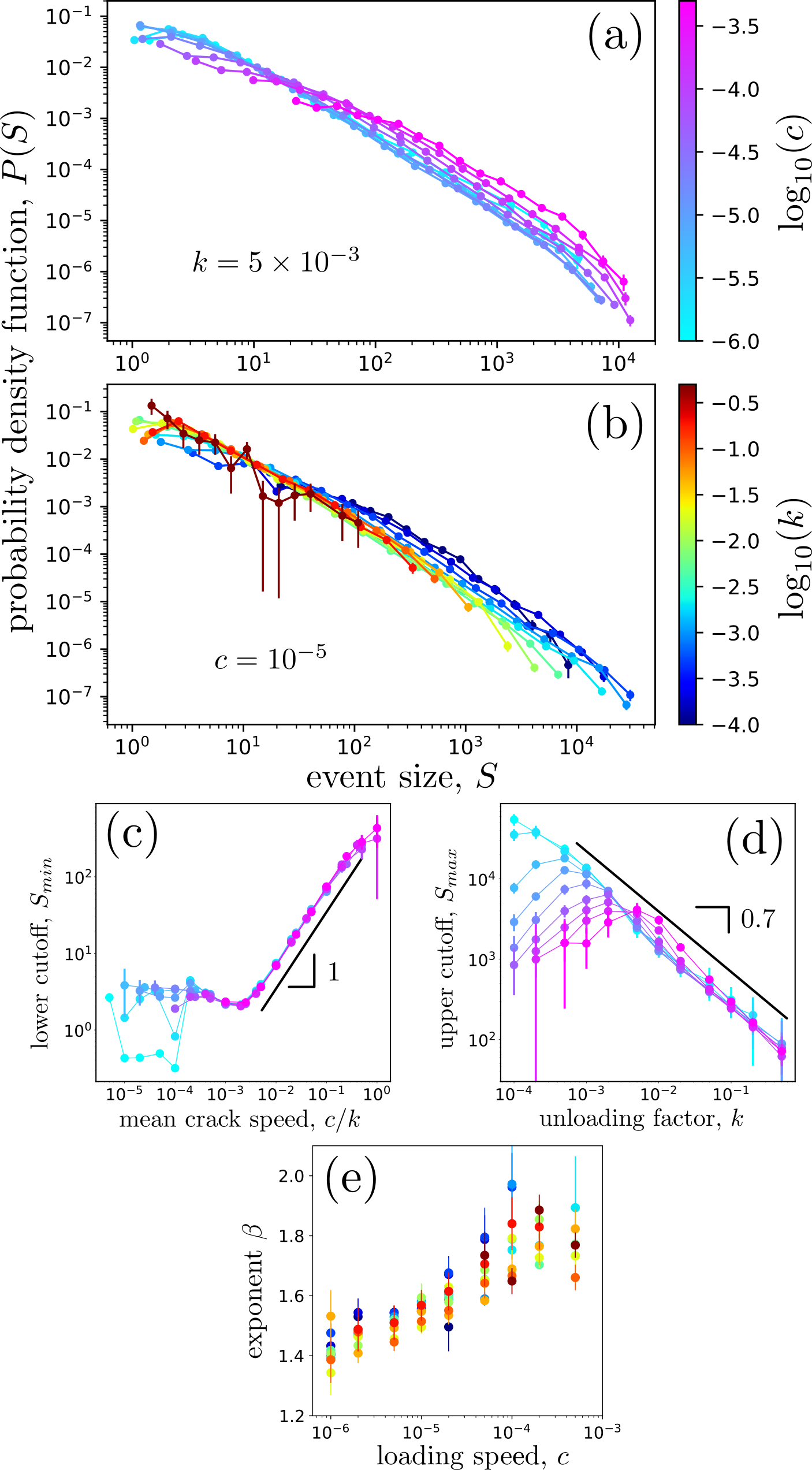}
\caption{(color online) Probability density function of the avalanche size $P(S)$ for different loading speed $c \in [10^{-6},5 \times 10^{-4}]$ keeping $k=5 \times 10^{-3}$ (panel a) and for different unloading factors $k \in [10^{-4},5 \times 10^{-1}]$ keeping $c=10^{-5}$ (panel b). $P(S)$ follows a power-law with exponent $\beta$ in between two cut-offs, $S_{min}$ and $S_{max}$. These three parameters are then determined by fitting each $P(S)$ curve using Eq. \ref{eqRG} and \ref{EqCutOff}. c: Evolution of the so-obtained lower cut-off $S_{min} = \frac{1}{\left< 1/S \right>}$ as a function of $\left< v \right>$. The dependency is almost linear; the black straight line has a slope of $1$. d: Evolution of the upper cut-off $S_{max} = \frac{\left< S^2 \right>}{\left< S \right>}$ as a function of $k$ for different values of $c$. $S_{max}$ decays as a power law with $k$, with a fitted exponent $\sim 0.7$ (black straight line). e: Exponent $\beta$ as a function of $c$ for different $k$ values. $\beta$ (very) slightly decreases with increasing $c$. On average, it remains close to $\sim 1.5$. On all panels, errorbars stand for $95$\% confident interval.}
\label{fig:RGck}
\end{figure}

We now turn to the role played by the control parameters, namely the (dimensionless) driving rate $c$ and unloading factor $k$ in Eq. \ref{eqLine}, onto the dynamics exhibited by the crack front. Figures \ref{fig:RGck}a and \ref{fig:RGck}b present the size distribution $P(S)$ obtained at different $k$ and $c$. Four observations emerge: 

\begin{itemize}
\item The lower cutoff $S_{min}$ increases with increasing $c$ and decreasing $k$; 
\item At fixed $c$, the upper cutoff $S_{max}$ displays a non-monotonic behavior with $k$. It first increases with $k$ at small $k$, reaches a maximum at $k^*$ and decreases at larger $k$; The increasing phase and the maximum position $k_*$ depend on $c$. Conversely, the decreasing phase seems independent of $c$. 
\item Over the whole range explored, $P(S)$ is in first approximation compatible with the gamma distribution (with lower cut-off) provided by Eq. \ref{eqRG};
\item the exponent $\beta$ (slope in the log-log representation) barely depends on $c$. 
\end{itemize} 

The lower and upper cutoffs of $P(S)$ are either measured directly by fitting the experimental curves with Eq. \ref{eqRG}, or by using:

\begin{equation}
\begin{split}
	S_{min} = 1/\left< 1/S \right> \\
	S_{max} = \left< S^2 \right>/\left< S \right>
\end{split}
\label{EqCutOff}
\end{equation}

\noindent It was checked that both definitions lead to the same results, but for a prefactor close to unit.

The lower cutoff is found to increase almost linearly with $\left< v \right>=c/k$ (see Fig.\ref{fig:RGck}c):

\begin{equation}
	S_{min}(c,k) \sim \left< v \right> 
	\label{Sminck}
\end{equation}

\noindent The saturation of $P(S)$ for $S \leq S_{min}$ may also be a consequence of the prescribed threshold $v_{th}=\left< v \right>$. Indeed, by setting a small and constant threshold $v_{th}$, in has been shown \cite{bares2014_ftp} that neither $c$ nor $k$ affect the value of $S_{min}$.   

The upper cutoff, $S_{max}$ displays a non-monotonic behavior with $k$. This behavior can be qualitatively understood in the framework of the depinning transition. At small velocity $\left< v \right>$, the quasi-static limit is reached and each burst corresponds to a single depinning avalanche. In this limit, the avalanche statistics is scale-free up to a correlation length $\xi_k\sim 1/\sqrt{k}$ \cite{bonamy2009_jpd}. When $\left< v \right>$ increases, a second velocity dependent length-scale is involved: 

\begin{equation}
\xi_v \sim \left< v \right>^{-\nu/\theta},
\end{equation}

\noindent with $\nu=1.625$ and $\theta=0.625$ \cite{bonamy2009_jpd,Duemmer2007_jsm}. The cutoff $S_{max}$ is governed by this length scale when $\xi_v \leq \xi_k$. The crossover between these two regimes occurs when $\xi_v \sim \xi_k$, that is:

\begin{equation}
k^* \sim c^{2\nu/(\theta+2\nu)}
\end{equation}

\begin{figure}
\centering \includegraphics[width=1.\columnwidth]{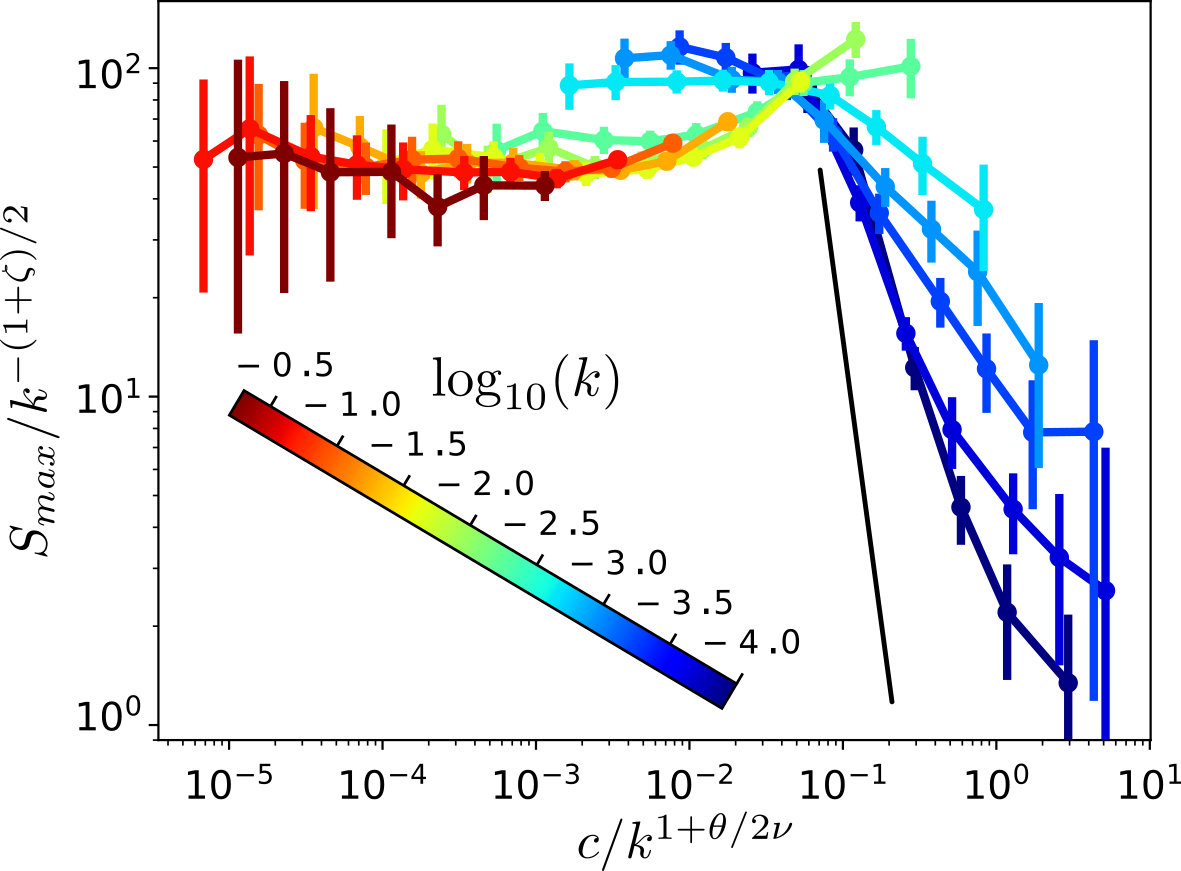}
\caption{(color online) Evolution of $S_{max}/k^{-(1+\zeta)/2}$ \textit{vs.} $c/k^{1+\theta/2\nu}$. A collapse of all curves is predicted within the depinning transition framework (Eq. \ref{Smaxcollapse}). This collapse is fulfilled for large $k$ and small $c$. Conversely, it is not fulfilled at large $c$ and/or small $c$. Tis departure results from size and discretisation effects. Straight black line indicates the power-law of exponent $-\nu(1+\zeta)/\theta \simeq -3.64$ predicted within the depinning transition framework.}
\label{fig:Smaxcollapse}
\end{figure}

In the framework of the depinning transition, $S_{max}$ is then expected to evolve with $c$ and $k$ as:

\begin{equation}
\begin{array} {l}
S_{max}(c,k)\sim k^{-(1+\zeta)/2} \times g(u=c/k^{1+\theta/2\nu}),\\
$with$~g(u) \sim \left\{
\begin{array}{l l}
1 & $if$ ~ u \ll 1  \\
u^{-\nu(1+\zeta)/\theta} & $if$ ~ u \gg 1
\end{array}
\right.
\end{array}
\label{Smaxcollapse}
\end{equation}

\noindent where the roughness exponent $\zeta = 0.4$ \cite{rosso2002_pre}. Note that this prediction holds in the continuum limit, when finite size and discretization effect can be neglected: $1 \ll \{\xi_k,\xi_v\} \ll L$. In Fig.  \ref{fig:RGck}d, we show the non-monotonic behavior of $S_{max}$ with $k$ and the agreement between the data and Eq. \ref{Smaxcollapse} for large $k$. To go deeper into the comparison, we looked at the variation of $S_{max}$ as a function of $c$ at fixed $k$. In Fig. \ref{fig:Smaxcollapse} shows $S_{max}/k^{(1+\zeta)/2}$ \textit{vs.} $c/c^*$ with $c^* \sim k^{1+\theta/2\nu}$. For small $c$ we found the collapse of the plateau consistent with the large scale $k$ behavior of Eq. \ref{Smaxcollapse}. For larger values of $c$, $S_{max}$ decreases with increasing $c$ as $\xi_v$ is dominant. The power-law predicted by Eq. \ref{Smaxcollapse} is shown by the plain black line and the agreement is not fulfilled. This departure results from size and discretization effects: at large $k$, $\xi_v$ starts being dominant only at short length-scales. At smaller $k$, $\xi_v$ is larger and the decay approaches the expected one but the system size is too small as can be seen from the non-collapse of the plateau.    

The distribution $P(S)$ is well fitted here by the gamma distribution provided in Eq. \ref{eqRG}. It is worth to note that, in the quasistatic limit ($\left< v \right> \rightarrow 0$ and subsequently $v_{th} \rightarrow 0$), $P(S)$ displays a stretched exponential behavior with exponents that can be computed by FRG techniques \cite{rosso09_prb}.

Within errorbars, $\beta$ is independent of $k$. Conversely, it increases slightly with $c$, from $\sim 1.4$ at $c = 10^{-6}$ to $\sim 1.6$ at $c = 10^{-4}$ (see Fig.\ref{fig:RGck}e). The value at vanishing $c$ is in agreement with the FRG value $\beta(c\rightarrow 0) =1.28 $ \cite{bonamy2009_jpd}. The larger value observed at finite $c$ may be an effect of the finite threshold, which, by dividing the depinning avalanches into smaller ones, could yield a larger effective exponent $\beta$ \cite{jagla2014_prl}. Indeed, similarly to what has already been discussed for $S_{min}$, making a different choice for the prescribed threshold $v_{th}$ (that is setting it to a constant prescribed low value ($v_{th}=10^{-3}$ as in \cite{bares2014_ftp}) yields a constant $\beta$ contrary to what is observed here. This emphasizes the importance of finite thresholding in the analysis of the selection of scales in crackling dynamics.     

\subsection{On the selection of waiting time law}\label{Sec3:WT}

\begin{figure}
\centering \includegraphics[width=1.\columnwidth]{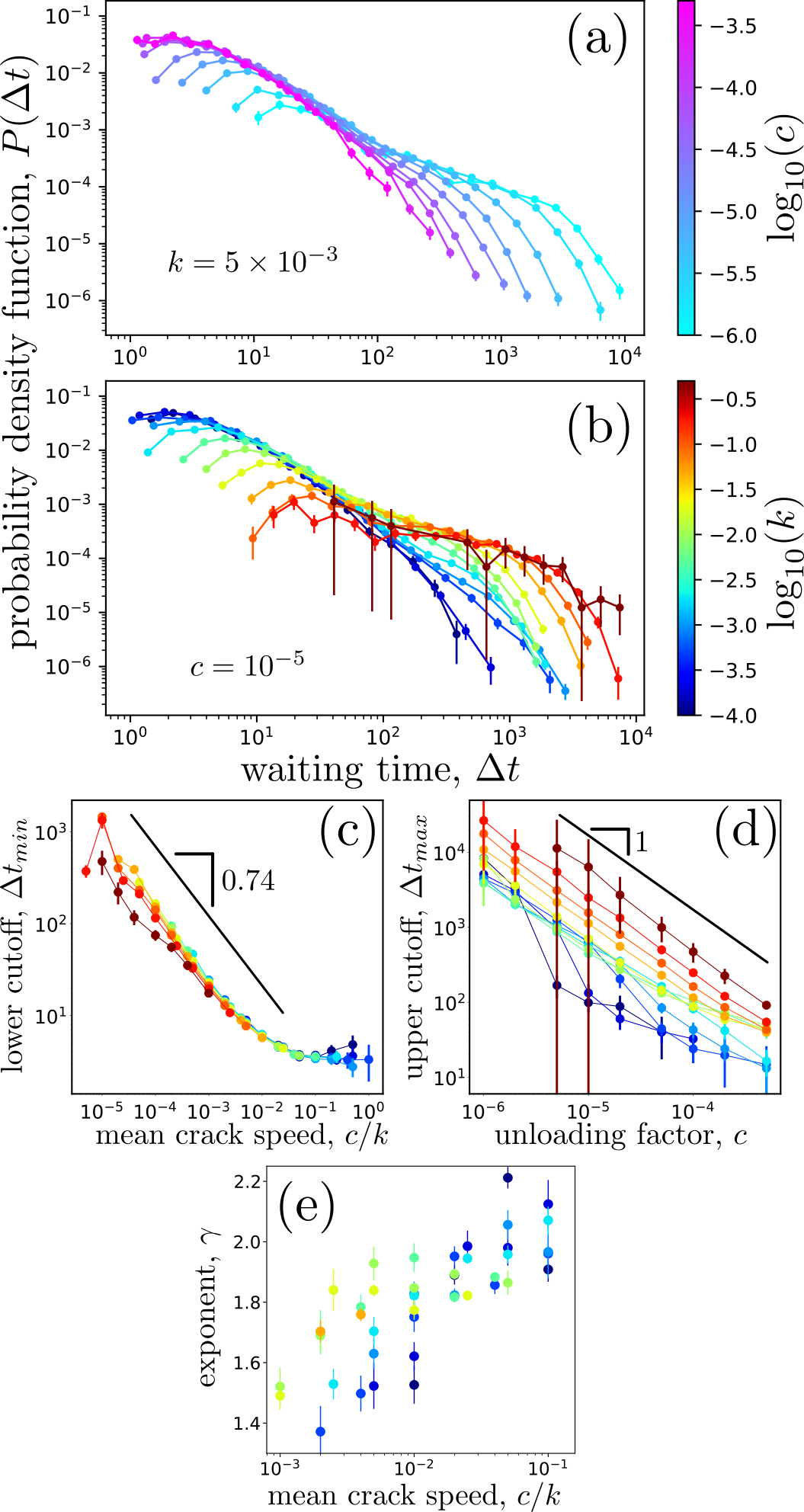}
\caption{(color online) a: Probability density function of the waiting time between two consecutive avalanche of size larger than $S_{th}=4$, $P(\Delta t)$. Curves are plotted for different loading speed $c \in [10^{-6},5 \times 10^{-4}]$ keeping $k=5 \times 10^{-3}$ (panel a) and for different unloading factors $k \in [10^{-4},5 \times 10^{-1}]$ keeping $c=10^{-5}$ (panel b). $P(\Delta t)$ follows a power-law with exponent $\gamma$ in between two cut-offs $\Delta t_{min}$ and $\Delta t_{max}$. These three parameters are then determined by fitting each $P(\Delta t)$ curve using Eq. \ref{eqWT} and equations equivalent to Eq. \ref{EqCutOff} for $\Delta t$. c: Evolution of $\Delta t_{min}$ as a function of the mean crack speed $\left< v \right>=c/k$. $\Delta t_{min}$ decays as a power-law with $\left< v \right>$, with a fitted exponent close to $0.74$ (black straight line). d: Evolution of $\Delta t_{max}$ as a function of $c$. $\Delta t_{max}$ decays as a power law with $c$, with a fitted exponent close to $1$ (black straight line). e: Exponent $\gamma$ as a function of $\overline{v}$. $\gamma$ increases logarithmically with $\overline{v}$ and goes from $1.4$ at $\overline{v}\simeq 10^{-3}$ to $2.2$ at $\overline{v}\simeq 10^{-1}$. The different colors in panels $c$ to $e$ correspond to different values $k$ according to the color bar provided in panel $e$. In all panels, the vertical bars stand for $95$\% error-bars.}
\label{fig:WTck} 
\end{figure}

Figure \ref{fig:WTck} synthesizes the effect of the parameters $c$ and $k$ onto the distribution of waiting time. The main effect observed here is that decreasing $c$ and/or increasing $k$ flatten the curve (in logarithmic axis), making the effective exponent $\gamma$ larger (see Figs.\ref{fig:WTck}a and \ref{fig:WTck}b); here again, $\left< v \right>=c/k$, seems to be the relevant parameter and $\gamma$ goes from $\sim 1.4$ to $\sim 2.2$ as $\overline{v}$ goes from $10^{-3}$ to $10^{-1}$ (see Fig.\ref{fig:WTck}e). The value at vanishing speed is close to $1.5$ which corresponds to the exponent of the power-law statistics of the avalanche duration in the quasi-static limit ($\alpha = 1+\zeta/\kappa \simeq 1.50$ where $\kappa=0.77$ is the dynamic exponent for the long range depinning transition \cite{Duemmer2007_jsm}). This scaling symmetry between the waiting time statistics and the avalanche duration statistics has indeed been invoked in \cite{janicevic2016_prl} when a finite threshold $v_{th}=\left< v \right>$ is prescribed. The increase of $\gamma$ with $\overline{v}$ is similar to what is observed experimentally, in \cite{bares2018_natcom}.    

In contrast to what has been observed for the size $S$ (Sec. \ref{Sec3:RG}), both the minimal and maximal waiting times $\Delta t_{min}$ and $\Delta t_{max}$ decreases with $\overline{c}$ (or $\left< v \right>$) (Fig.\ref{fig:WTck}c and d). This can be understood if one thinks that the nucleation rate of new avalanches is proportional to $c$. Hence, the typical waiting time, $\widetilde{\Delta t}$ between successive avalanches goes as $1/c$. Indeed, as long as the duration of the avalanche is negligible, in order to nucleate a new avalanche, one should increase the force $\delta F =c \delta t$ by a fixed amount $\sim 1/L$ \footnote{This is generic to the depinning transition where the probability density function of the distances from instability threshold does not vanish at origin. Then the most instable among $L$ elementary blocks always scales as $1/L$ \cite{lin2014scaling}}. 

This scaling is perfectly obeyed by $\Delta t_{max}$ for large $k$ and small $c$. When $k$ decreases, avalanche duration becomes larger. This induces a decrease of the measured $\Delta t_{max}$, which does not coincide exactly with the time interval between successive nucleation events anymore. In this regime, the $1/c$ scaling is only an upper bound for $\Delta t_{max}$ that is shifted all the more so as $k$ decreases. This regimes survives as long as the avalanche duration remains small with respect to $\widetilde{\Delta t}$. As the upper cutoff is mainly limited by $\xi_k$ (see Sec. \ref{Sec3:RG}), this avalanche duration is expected to increase with decreasing $k$ and, for small enough $k$ to become of the order of $\widetilde{\Delta t}$. At this point, the depinning avalanches coalesce together and the waiting time in between drops abruptly. In this coalescence regime, it is the finite threshold value ($c/k$) that controls $\Delta t_{max}$.   

\subsection{On the conditions leading to seismic-like organization}\label{Sec3:phasediagram}

\begin{figure}
\centering \includegraphics[width=1\columnwidth]{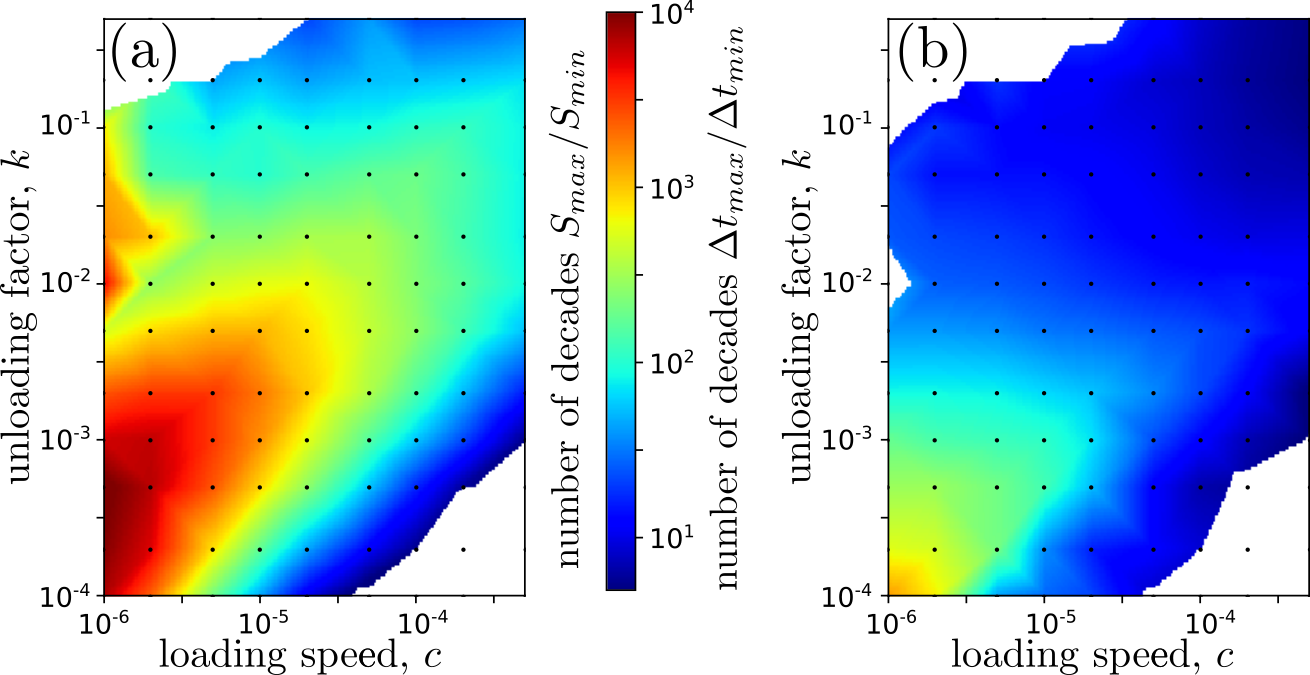}
\caption{(color online) Phase diagram showing the $\{c,k\}$ conditions to observe : (a) crackling dynamics, that is scale-free statistics for size over a significant number of decades; and (b) temporal seismic-like intermittency, that is a scale-free statistics for interevent time over a significant number of decades. In both maps, the $c$ and $k$ axes are logarithmic. The color indicates $\log_{10}(S_{max}/S_{min})$ (panel a) and $\log_{10}(\Delta t_{max}/\Delta t_{min})$ (panel b) according to the colorbar shown in between the two panels. The range of parameters  $\{c,k\}$ allowing to observe extended scale-free statistics for size is much larger than that required to observe extended scale-free temporal correlation (red areas in panels a and b).}
\label{fig:Diag}
\end{figure}

Finally, to unravel the conditions favoring seismic-like behavior, that is a scale free statistics of size {\em and} waiting time, we plotted, in Fig.\ref{fig:Diag}, the number of decades over which a scale free statistics is observed for both quantities. 

Concerning the sizes, two zones with many decades of scale-free statistics are observed (Fig. \ref{fig:Diag}a): A first, fairly large, one in the left-handed/lower part of the diagram (small $k$, small $c$) and second smaller one at the left-handed/upper part (finite $k$, small $c$). The fact that $c$ should be small is well understood: small $c$  yields small $\left< v \right>$, which  favors both large $S_{max}$ and small $S_{min}$ (see Fig. \ref{fig:RGck}c and d). Conversely, $k$ has two antagonist effects: Increasing $k$ makes $\xi_k$ small, hence preventing large $S_{max}$; but at the same time, it makes $\left< v \right>$ small, yielding small $S_{min}$. The existence of the small zone with scale-free statistics at moderate $k$ and small $c$ is a consequence of this small $S_{min}$; it cannot be understood within the depinning theory but is a direct consequence of the experimental choice of a finite threshold equal $\left< v \right>$. 

Concerning the time clustering at the origin of the dynamics intermittency and of the fundamental seismic laws (see Sec. \ref{Sec2}), the scale free statistics is observed only in a tiny region with both small $k$ and $c$. Small $c$ is needed to observe large $S_{max}$ (Fig. \ref{fig:WTck}d) and small $k$ is needed to get large $\left< v \right>$, and subsequently small $S_{min}$. Note that the extension of the $\{c,k\}$ domain which allows observing scale free inter-event times over a significant range of scale is much smaller than that required for observing scale-free sizes. This explains why time clustering and seismic-like organization of avalanches sequences are barely reported in the context of depinning interfaces.

\section{Concluding discussion}\label{Sec4}

We analyzed here crackling dynamics exhibited by a long-range elastic 1D interface driven in a random potential. A slow and constant loading rate, $c$, is imposed and a finite unloading factor, $k$, is considered. As a result, the force applying onto the interface self-adjusts around the depinning threshold and the motion exhibit a steady avalanche dynamics, with a speed signal $v(t)$ fluctuating highly around an average value $\left< v \right>=c/k$. The avalanches were identified with the  bursts above this mean value, and their size and occurrence time were collected in event catalogs.    

The analysis of these catalogs revealed a statistical organisation similar to that reported in sismology: Both the avalanche size and inter-event time are power-law distributed. Moreover, the events form aftershocks sequences obeying the fundamental laws of seismology: Productivity law with a mean number of produced aftershocks scaling as a power-law with the mainshock size,  B\r{a}th's law with a ratio between the size of the mainshock and that of its largest aftershock is constant, and Omori-Utsu law with an afershock productivity rate decaying as a power-law with time. As experimentally observed in \cite{bares2018_natcom}, these laws do not reflect some non trivial correlation between size and occurrence time: They directly emerge from the scale-free statistics of energy (for the productivity and B\r{a}th's laws) and from that of inter-event time (for Omori's laws). 

The value of the loading rate and unloading factor has a drastic effect on the scaling exponents associated with the scale-free statistics of size and interevent time on one hand, and on the lower and upper cutoff limiting the scale-free regime on the other hand. The framework of the depinning transition allows understanding some of this effect; the dependency of $S_{max}$ with $k$ and that of $\Delta t_{max}$ with $c$ in particular. Still, this framework presupposes a quasi-static dynamics ($c \rightarrow 0$). A finite driving rate \eg requires us to work with a finite thresholding, which is shown here to have a drastic effect on the selection of $S_{min}$ and $\Delta t_{min}$. This finite thresholding has also been invoked to be responsible for the scale-free statistics of inter-event times \cite{janicevic2016_prl}. By making the depinning avalanches overlap partially, a finite driving rate also affect the effective values of the scaling exponents for size and interevent time \cite{stojanova14_prl}. Note finally that the dependencies of the lower and upper cutoffs with loading rate and unloading factor make it non-trivial to predict when crackling (scale-free size statistics) and/or seismic-like (scale free statistics for both size and interevent time) are observed. Small values for both $c$ and $k$ are required for the latter, while small $c$ and even moderate $k$ permits to observe crackling.


Beyond fracture problems, the universality class of long-range interface depinning also encompasses a variety of other physical, biological and social systems. The new insights obtained here on the time-size organization of fracture events and its evolution with loading rate and unloading factor likely extends to the other systems belonging to the same universality class. As a prospective work, the system size $N$, the random noise amplitude $\tilde{\eta}$ and the kernel nature and range of interaction is also likely to have a high influence on the time dynamics of this process.

\section*{Acknowledgments}

Support through the ANR project MEPHYSTAR is gratefully acknowledged.


\bibliographystyle{apsrev4-1}
\bibliography{biblio}

\end{document}